\documentclass[10pt, onecolumn, conference]{IEEEtran}

\usepackage[T1]{fontenc}
\usepackage[utf8]{inputenc}
\usepackage{babel}  
\usepackage[hyphens]{url}
\usepackage[shortlabels]{enumitem}
\usepackage[table,xcdraw]{xcolor}  
\usepackage{adjustbox}
\usepackage{tabularx}
\usepackage{booktabs}
\usepackage{algorithm}
\usepackage{algpseudocode}
\usepackage{makecell}
\usepackage{multicol}
\usepackage{amsmath}
\usepackage{amssymb}
\usepackage{amsfonts}
\usepackage{comment}
\usepackage{float}
\usepackage{footmisc}  
\usepackage{geometry}
\usepackage{graphicx}  
\usepackage{subcaption}  
\usepackage{underscore}
\usepackage{listings}
\usepackage{lmodern}
\usepackage{multirow}
\usepackage{scalefnt}
\usepackage{setspace}
\usepackage{xspace}
\usepackage{cmap}
\usepackage{svg}
\usepackage{threeparttable}
\newcommand\rurl[1]{\href{http://#1}{\nolinkurl{#1}}}

\newcommand{\omite}[1]{}

\newcommand{\orcidd}[1]{\href{https://orcid.org/#1}{\includegraphics[scale=0.2]{Orcid_icon.png}\hspace{2mm}}}

\newcommand*{\missingreference}{\colorbox{red}{?reference?}}
\newcommand*{\missingcitation}{\colorbox{red}{?citation?}}
\makeatletter
\def\@setref#1#2#3{%
  \ifx#1\relax
   \protect\G@refundefinedtrue
   \nfss@text{\reset@font\missingreference}%
   \@latex@warning{Reference `#3' on page \thepage \space
             undefined}%
  \else
   \expandafter#2#1\null
  \fi}
\def\@citex[#1]#2{\leavevmode
  \let\@citea\@empty
  \@cite{\@for\@citeb:=#2\do
    {\@citea\def\@citea{,\penalty\@m\ }%
     \edef\@citeb{\expandafter\@firstofone\@citeb\@empty}%
     \if@filesw\immediate\write\@auxout{\string\citation{\@citeb}}\fi
     \@ifundefined{b@\@citeb}{\hbox{\reset@font\missingcitation}%
       \G@refundefinedtrue
       \@latex@warning
         {Citation `\@citeb' on page \thepage \space undefined}}%
       {\@cite@ofmt{\csname b@\@citeb\endcsname}}}}{#1}}
\makeatother 

\title{MalDataGen: A Modular Framework for Synthetic Tabular Data Generation in Malware Detection} 

\author{
\IEEEauthorblockN{
Kayuã Oleques Paim,
Angelo Gaspar Diniz Nogueira\\
Diego Kreutz,
Weverton Cordeiro, Rodrigo Brandão Mansilha\\\vspace{1mm}
}

\IEEEauthorblockA{
LEA and PPGES, Universidade Federal do Pampa (UNIPAMPA), Brazil
}

\IEEEauthorblockA{
Universidade Federal do Rio Grande do Sul (UFRGS), Brazil
}
}

\begin{document}

\maketitle

\begin{abstract}
High-quality data scarcity hinders malware detection, limiting ML performance. We introduce MalDataGen, an open-source modular framework for generating high-fidelity synthetic tabular data using modular deep learning models (e.g., WGAN-GP, VQ-VAE). Evaluated via dual validation (TR-TS/TS-TR), seven classifiers, and utility metrics, MalDataGen outperforms benchmarks like SDV while preserving data utility. Its flexible design enables seamless integration into detection pipelines, offering a practical solution for cybersecurity applications.
\end{abstract}


\section{Introduction}
 
Modern machine learning algorithms, particularly deep learning architectures, depend on large-scale datasets with reliable annotations to achieve optimal performance. However, current methods for dataset collection and labeling require substantial resources and time investments \cite{zha2025data}. The field faces persistent challenges with data availability, as approximately 80\% of AI project failures stem from insufficient data quantity or quality \cite{ai2023top}.

These challenges are especially observable in domains with limited or imperfect data sources. Synthetic data generation offers a potential solution \cite{figueira2022survey,lee2025synthetic,hao2024synthetic}, creating artificial samples that maintain key characteristics of real-world data. In cybersecurity, this approach has been applied to improve malware detection systems \cite{peppes2023malware}, identify anomalous network traffic \cite{kumar2023synthetic}, and generate polymorphic malware variants \cite{dunmore2023comprehensive}.

We observe the development of several libraries and frameworks that help streamline and standardize the process for synthetic data generation. We provide a comparative overview of libraries for synthetic tabular data generation in Table~\ref{tab_related_work}. The table highlights the models and algorithms they use for generating synthetic data.

\begin{table*}[!htp]
\centering
\caption{Deep Learning-Based Libraries for Synthetic Tabular Data Generation.}
\label{tab_related_work}
\scriptsize
\renewcommand{\arraystretch}{1.4}
\begin{tabular}{>{\raggedright\arraybackslash}p{3cm} >{\raggedright\arraybackslash}p{1.7cm} >{\raggedright\arraybackslash}p{9cm}}
\toprule
\textbf{Library} & \textbf{Algorithms} & \textbf{Models} \\
\midrule
\textbf{Gretel Synthetics}\footnotemark[1] & 
GAN & 
DGAN, DPGAN, ACTGAN \\
\midrule
\textbf{YData}\footnotemark[2] & 
GAN & 
GAN, cGAN, WGAN, WGAN-GP, DRAGAN, CramerGAN, CWGAN-GP, CTGAN   \\ 
\midrule
\multirow{3}{*}{\textbf{SDV}\footnotemark[3]} 
&  Statistical  & Copula \\ 
& GAN & CTGAN \\
& AE & TVAE \\
\midrule
\multirow{2}{*}{\textbf{MalDataGen}\footnotemark[4]} 
& GAN & GAN, WGAN-GP, cGAN \\
\multirow{2}{*}{\textbf{(This Work)}} & AE & AE, VAE, VQ-VAE \\ 
& Diffusion & LDM \\
\bottomrule
\end{tabular}
\end{table*}

\footnotetext[1]{\url{https://gretel.ai/}}
\footnotetext[2]{\url{https://ydata.ai/}}
\footnotetext[3]{\url{https://docs.sdv.dev/sdv}}
\footnotetext[4]{\url{https://github.com/SBSeg25/MalDataGen}}

Existing libraries face two main constraints: limited flexibility for custom modifications and a narrow range of pre-implemented algorithms. We address these issues with a new Python-based modular and extensible framework, named MalDataGen\footnotemark[4], with broader algorithms support. While Table~\ref{tab_related_work} shows most current tools focus on GAN-based approaches \cite{mirza2014conditional}, and SDV includes Autoencoders \cite{kingma2013auto}, our solution expands on these foundations. We incorporate additional implementations of established methods and present what we believe are the first tabular data applications of VQ-VAEs \cite{van2017neural} and Latent Diffusion Models \cite{Rombach2022}. Our solution is also designed to be composable, enabling scientists and practitioners to assemble other models from our basic components.

Our evaluation shows that MalDataGen outperforms SDV, which currently offers the widest range of algorithms. We assess all generative models from both libraries using seven classifiers and follow the methodology from \cite{esteban2017real}, implementing two validation approaches: TR-TS (Train on Real - Test on Synthetic) and TS-TR (Train on Synthetic - Test on Real).

The paper is organized as follows. Section~\ref{sec_SynDataGen} describes our framework. Section~\ref{sec_experiments_results} presents and discusses the results, with concluding remarks in Section~\ref{sec_conclusion}.

\section{MalDataGen: Composable Generative Modeling Framework}
\label{sec_SynDataGen}

\begin{figure}[!hbt]
    \centering
    \includegraphics[width=1\linewidth]{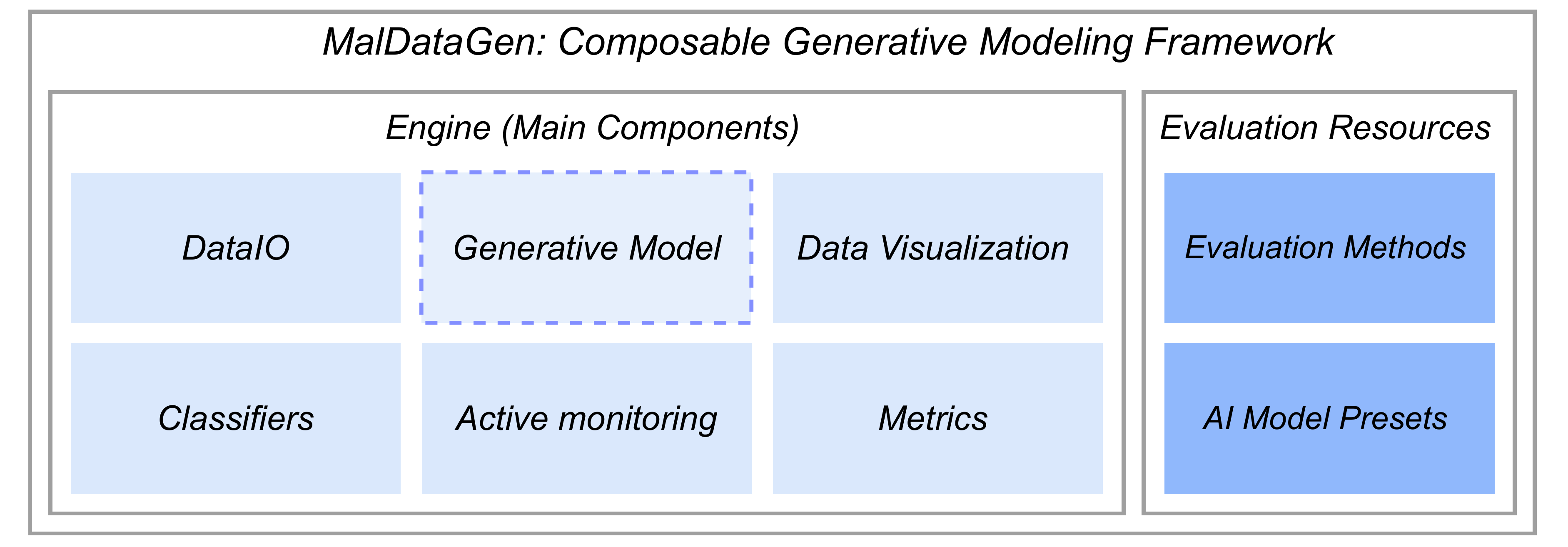}
    \caption{Overview of the MalDataGen composable framework.}
    \label{fig_arch}
\end{figure}

Figure~\ref{fig_arch} shows the architecture of \textit{MalDataGen}. The design includes two core components: (1) the \textit{Engine} with components for developing and managing deep learning-based generative models, and (2) the \textit{Evaluation Resources} for validating synthetic data quality. 

\subsection{Engine}
\label{sec_synthetic_ocean}

The \textit{Engine} contains modules for development, training, and orchestration of deep learning-based generative models. We organize the library into six key modules: \textit{DataIO}, \textit{Data Visualization}, \textit{Classifiers}, \textit{Metrics}, \textit{Active Monitoring}, and \textit{Generative Models}. Additionally, the Engine includes auxiliary modules, such as those for handling arguments and exceptions. This structure helps manage complexity while allowing customization and reuse.

We built the \textit{DataIO module} to handle data acquisition, transformation, and storage. It works with structured formats like \textit{.csv} and \textit{.xls}, supporting dataset normalization, schema-aware editing, and format-preserving serialization. The pluggable I/O layer adapts to new file formats and domain-specific representations.

The \textit{Data Visualization module} offers tools for analyzing both real and synthetic datasets. Its capabilities include visualization of statistical dependencies through correlation heatmaps, cluster analysis using techniques like K-Means with UMAP, and graphical representation of performance metrics such as confusion matrices and bar plots. These components support customization of visual styles and export options.

Our \textit{Classifiers module} contains 15 supervised learning algorithms, including SVM, Random Forest, and MLP. The abstract interface design enables efficient comparison across different modeling approaches.

In the \textit{Metrics module}, we implement measures for predictive performance (F1-score, AUC, MCC) and distribution similarity (Jensen-Shannon Divergence, Wasserstein Distance, MMD). These help evaluate how well synthetic data matches real data distributions.

The \textit{Active Monitoring module} oversees the entire data pipeline. It detects faults like NaN values and divergence, tracks resources including memory usage, and maintains time-stamped logs. The system includes convergence-aware callbacks for runtime adjustments.

For the \textit{Generative Models} module, we provide a suite of fully configurable synthetic-data generators that allow architecture customization, hyperparameter tuning, and the selection of diverse generation strategies. Out-of-the-box, the module supports CTGAN, variational autoencoders (VAEs), and Gaussian copula models, each capable of producing samples that faithfully preserve the statistical properties of the original dataset.

Crucially, we have extended each base implementation to better suit Android‑malware generation. In our GAN and VAE variants, we introduce an embedding layer for malware class labels (benign vs. malicious) and a projection layer to streamline feature preprocessing. For the latent diffusion model (LDM), we depart from the original TabDDPM design~\cite{kotelnikov2023tabddpm} by integrating a VAE subnetwork: this ensures a continuous, lower‑dimensional latent space. Finally, during the reverse‑diffusion (denoising) stage, we employ a bespoke U‑Net composed of multi‑channel feedforward layers and residual temporal‑encoding blocks, and train it end‑to‑end with a combined KL‑divergence plus MSE loss.

\begin{figure}[!htb]
\centering
\includegraphics[width=1\linewidth]{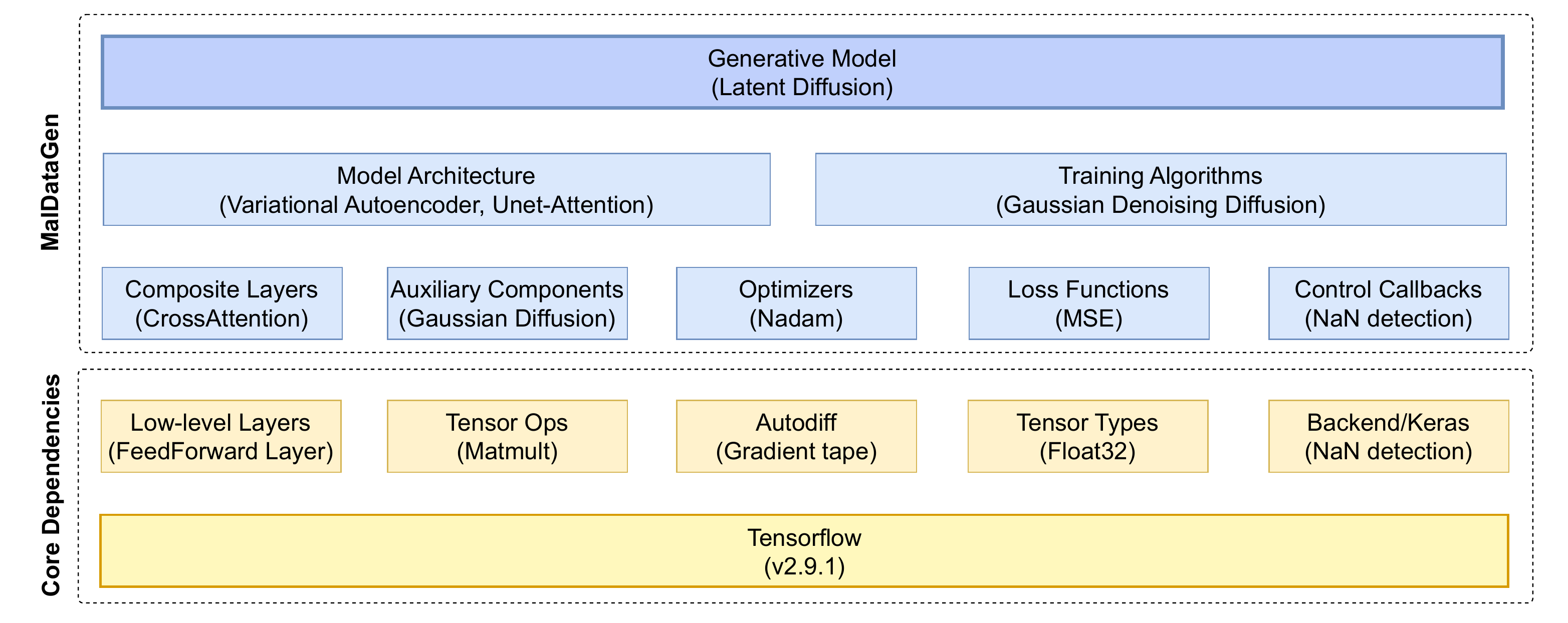}
\caption{Overview of the \textit{Generative Models} architecture (with illustrative examples in parentheses).}
\label{fig_architecture}
\end{figure}

The \textit{Generative Models module} follows a two-layer architecture shown in Figure~\ref{fig_architecture}. We organize the upper layer with functional building blocks, while the lower layer handles core computational dependencies including TensorFlow and essential Python libraries.

For the functional building blocks, we implement reusable components that support flexible pipeline construction. The \textit{optimizers} includes gradient-based algorithms like Adam, Nadam, and SGD, which we can configure for different learning scenarios. We provide various \textit{loss functions} such as MSE, Cross-Entropy, and KL Divergence to handle different training objectives. The \textit{composite layers} incorporates abstractions like Time Embedding and Sampling Layers that work with architectures such as VQ-VAEs. Additional \textit{auxiliary components} includes Gaussian diffusion processes and attention mechanisms. For training supervision, we implement \textit{control callbacks} that handles early stopping and resource monitoring.

The training architecture separates learning dynamics from model structure. We implement multiple \textit{training algorithms} including GANs, VAEs, and Diffusion Models that work with different \textit{model architectures} like UNet and Encoder-Decoder pairs. This separation allows combining architectures with different training objectives without structural modifications. For example, a UNet backbone can work with either VAE objectives or DDPM training.

At the foundation, we build on TensorFlow's computational framework\footnote{\url{https://www.tensorflow.org}} for core operations. The \textit{low-level layers} provide basic neural network components including convolutional and dense layers. We implement essential \textit{tensor operations} like matrix multiplication and other algebraic routines. The automatic differentiation system enables gradient computation through reverse-mode differentiation. For tensor management, we include shape and data type specifications through \textit{tensor typing}. The system integrates with Keras\footnote{\url{https://keras.io/api/}} for execution management and high-level API access.

This architecture combines TensorFlow's computational capabilities with modular components for generative modeling. The layered approach maintains performance while supporting flexible configuration and extension of the core functionality. Each component interacts through well-defined interfaces, allowing researchers to modify or extend specific aspects without affecting other system parts.

We augment our discussion with a comprehensive set of visuals. First, five in-depth figures unpack the internal workings of each generative model, clearly illustrating their key architectural distinctions. Complementing these, eight Mermaid diagrams chart the entirety of the MalDataGen framework -- depicting everything from the high-level system layout and object-oriented class relationships to the complete data-processing pipeline, evaluation routines, training workflows, and metrics architecture. Together, these graphical resources present a unified, detailed perspective on how MalDataGen’s modules interlock to produce and rigorously assess high-quality synthetic data for cybersecurity applications. For the full collection of diagrams and detailed explanations, please visit our GitHub repository\footnote{\url{https://github.com/SBSeg25/MalDataGen}}.

\subsection{Evaluation Resources}
\label{sec_evaluation_resources}

Our \textit{evaluation resources} provides systematic protocols for assessing synthetic data quality and usefulness. These resources consist of two core components: the \textit{Evaluation Methods} and \textit{AI Model Presets} modules.

The \textit{Evaluation Methods} component implements validation approaches including $k$-fold cross-validation, where we partition datasets into $k$ subsets for iterative training and testing. We also employ domain transfer evaluation through two complementary approaches: \textit{Train-on-Real/Test-on-Synthetic} (TR-TS) and \textit{Train-on-Synthetic/Test-on-Real} (TS-TR). These methods allow us to assess both data transferability and model robustness using paired classifiers from our classification tools.

For the \textit{AI Model Presets}, we maintain version-controlled configurations containing optimized parameters such as learning rates for GAN components and training epoch counts. The module includes architectural templates for supported network types and initialization parameters for models from our generative components. These presets help ensure consistent evaluation across different experimental setups.

The resources incorporate standardized metrics to compare synthetic and original datasets across multiple dimensions. All components integrate with other library modules through defined interfaces, supporting reproducible assessment of synthetic data quality. We designed the system to balance comprehensive evaluation with practical usability, allowing researchers to focus on their specific validation needs.

\section{Results and Discussion}
\label{sec_experiments_results}

We present and analyze our key findings using the SVM (Support Vector Machine) classifier to demonstrate patterns observed across classifiers. Complete results for all classifiers and hyperparameter configurations are available in our public repository.

For our experiments, we employed the Androcrawl dataset from the MalwareDataHunter Project’s public repository\footnote{\url{https://github.com/Malware-Hunter/datasets.git}}. This dataset comprises 20,340 samples—10,170 malware and 10,170 benign—each described by 136 features. Prior to training, we applied chi-square feature selection to retain the top 200 features, then down-sampled each class to 10,000 instances (20,000 total) where needed to ensure balance.

We leave a comprehensive examination of synthetic data’s influence on class balance—and a rigorous investigation into potential data leakage stemming from our feature-selection process—for subsequent studies.

\begin{figure}[!htp]
\centering
\includegraphics[width=1\linewidth]{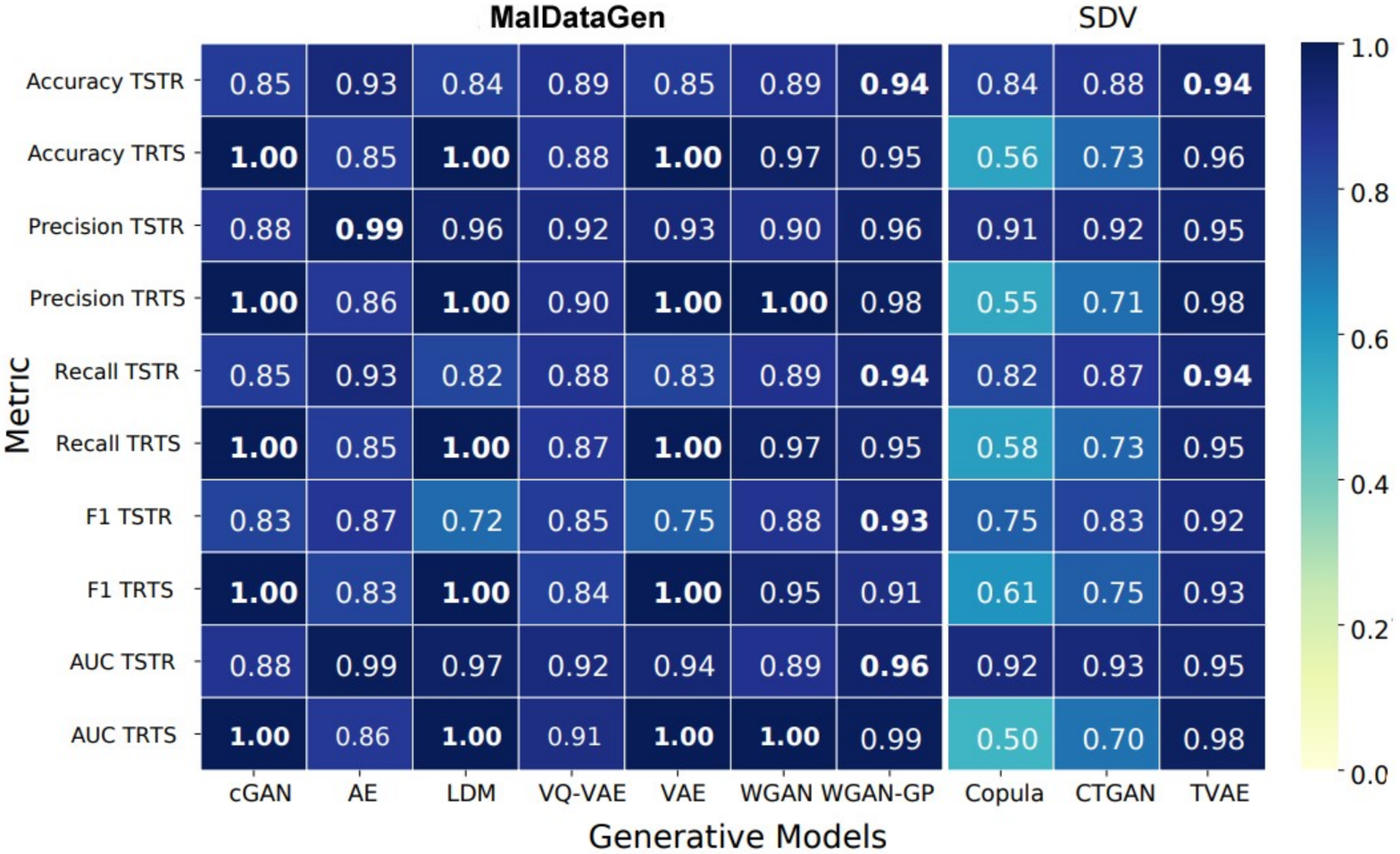}
\caption{Utility assessment: Binary classification metrics for SVM classifier performance using data generated by different models.}
\label{fig_comparisson_models_svm}
\end{figure}

Figure \ref{fig_comparisson_models_svm} presents a heatmap of average 5‑fold cross‑validation scores for SVM classifiers trained on synthetic samples generated by both our MalDataGen framework and the SDV library. Darker cells indicate stronger performance (closer to 1.0) and lighter cells indicate weaker performance (closer to 0.0). We report utility metrics—accuracy, precision, recall, F1‑score, and AUC—with paired rows contrasting the TS‑TR and TR‑TS evaluation scenarios.

Our results reveal notable performance variations across generative models. The WGAN-GP and WGAN implementations show strong performance across all metrics, often achieving near-perfect scores in both evaluation scenarios. CGAN and AE follow closely, with CGAN particularly excelling in TR-TS metrics. VAE and LDM maintain high performance but show slightly lower TS-TR F1-scores due to reduced recall. Our VQ-VAE implementation, while the lowest-performing among our models, maintains all metrics above 0.84.

SDV's models demonstrate weaker TR-TS performance. The Copula model shows limited efficacy, with critical TR-TS metrics approaching random classification levels. CTGAN shows improved TR-TS capability over Copula but remains behind our models. SDV's TVAE performs exceptionally, matching our top-performing WGAN-GP across both evaluation paradigms.

These results suggest that with proper hyperparameter optimization, our models can match or surpass existing open-source implementations like SDV. The performance differences highlight the importance of model selection and tuning for synthetic data generation tasks.

We present the mean distance metric results from our 5-fold evaluation in Table~\ref{tab_fidelitym}. The table shows distance measurements between synthetic and real data, where lower values indicate higher fidelity. Each column represents a different generative model from our framework and the SDV library.

\begin{table}[!htb]
\caption{Fidelity assessment: Distance metrics comparing real data to synthetic outputs across generative models (lower values indicate better alignment).}
\label{tab_fidelitym}
\centering
\resizebox{\columnwidth}{!}{%
\renewcommand{\arraystretch}{1.5}
\begin{tabular}{c|ccccccc|ccc}
\hline
\multirow{2}{*}{Metrics} & \multicolumn{10}{c}{Models} \\ 
\cline{2-11}
 & \multicolumn{7}{c|}{Ours} & \multicolumn{3}{c}{SDV} \\ 
\cline{2-11}
 & cGAN & AE & LDM & VQ-VAE & VAE & WGAN & WGAN-GP & Copula & CTGAN & TVAE \\ 
\hline
Euclidean Distance $\downarrow$ & 3.59 & 3.97 & 3.46 & 4.62 & 3.46 & \textbf{3.45} & 3.89 & 4.43 & 4.51 & 4.29 \\ 
\hline
Hellinger Distance $\downarrow$ & 166.08 & 181.03 & 160.57 & 210.20 & 160.88 & \textbf{159.50} & 179.00 & 200.90 & 204.74 & 195.28 \\ 
\hline
Manhattan Distance $\downarrow$ & 405.7 & 483.39 & 379.17 & 650.19 & 380.63 & \textbf{374.16} & 471.21 & 593.54 & 616.59 & 560.79 \\ 
\hline
Hamming Distance $\downarrow$ & 2.98 & 3.55 & 2.79 & 4.78 & 2.80 & \textbf{2.75} & 3.46 & 4.36 & 4.53 & 4.12 \\ 
\hline
Jaccard Distance $\downarrow$ & 0.65 & 0.66 & \textbf{0.60} & 0.74 & \textbf{0.60} & 0.61 & 0.66 & 0.70 & 0.71 & 0.69 \\ 
\hline
\end{tabular}%
}
\end{table}
The results show a consistent pattern with our utility metric findings. Our WGAN implementation achieves the lowest distance measures across most metrics, indicating the closest similarity to real data. The WGAN-GP, cGAN, and AE models follow with slightly higher but still competitive distance values. Among our implementations, VQ VAE shows the largest distance from real data distributions.

For SDV's models, we observe generally higher distance metrics, with Copula and CTGAN showing the greatest dissimilarity to real data. This aligns with their weaker performance in utility evaluations. SDV's TVAE stands as an exception, matching our WGAN-GP in both fidelity and utility metrics, demonstrating comparable effectiveness in generating authentic synthetic data.

The fidelity assessment reinforces the utility metric results, showing that models producing data closer to the real distribution also perform better in downstream tasks. The consistent performance across both evaluation dimensions suggests that careful model selection and optimization can yield synthetic data that preserves both statistical properties and practical usefulness.
 
\section{Final Considerations}
\label{sec_conclusion}

\noindent \textbf{Demonstration.}
We will demonstrate how MalDataGen operates through a practical example executed on one of our devices. This demonstration will highlight its configuration parameters, execution pipeline  and an analysis of the produced such as heatmaps, confusion matrices, and training curves. Examples of these outputs are presented in Appendix A.

\noindent \textbf{Conclusion.}
We presented \textit{MalDataGen}, a \textit{Composable Generative Modeling Framework} designed for the generation of synthetic tabular data, with a specific focus on cybersecurity applications. Our implementation showed comparable or superior results to \textit{SDV} in the evaluated scenarios. It also introduced a modular architecture that supports extensibility, a set of pre-configured generative models, and a methodology based on two validation strategies (TR-TS and TS-TR), which helped assess the quality of the generated data. We used visualization methods to support exploratory analysis and to examine the qualitative characteristics of the synthetic data.

\noindent \textbf{Future directions.}
We plan to improve the library by adding new generative models, classifiers, and metrics. We also expect to integrate it with tools for data analysis and generation to increase interoperability. Our evaluation resources will be expanded to include other libraries, such as \textit{YData} and \textit{GretelSynthetics}, along with additional datasets and use cases.




\section*{Acknowledgements}

The research was supported by RNP (Hackers do Bem Program – GT Malware DataLab), CAPES (Financing Code 001), the FAPERGS via calls 02/2022, 08/2023, and 09/2023 (grant agreements 24/2551-0001368-7 and 24/2551-0000726-1), and by FAPESP (processes 2020/05183-0 and 2023/00816-2).

\bibliographystyle{IEEEtran}
\bibliography{references}
 
\newpage

\section*{Examples of outputs}

\begin{figure}[!htb]
    \centering
    \includegraphics[width=1\linewidth]{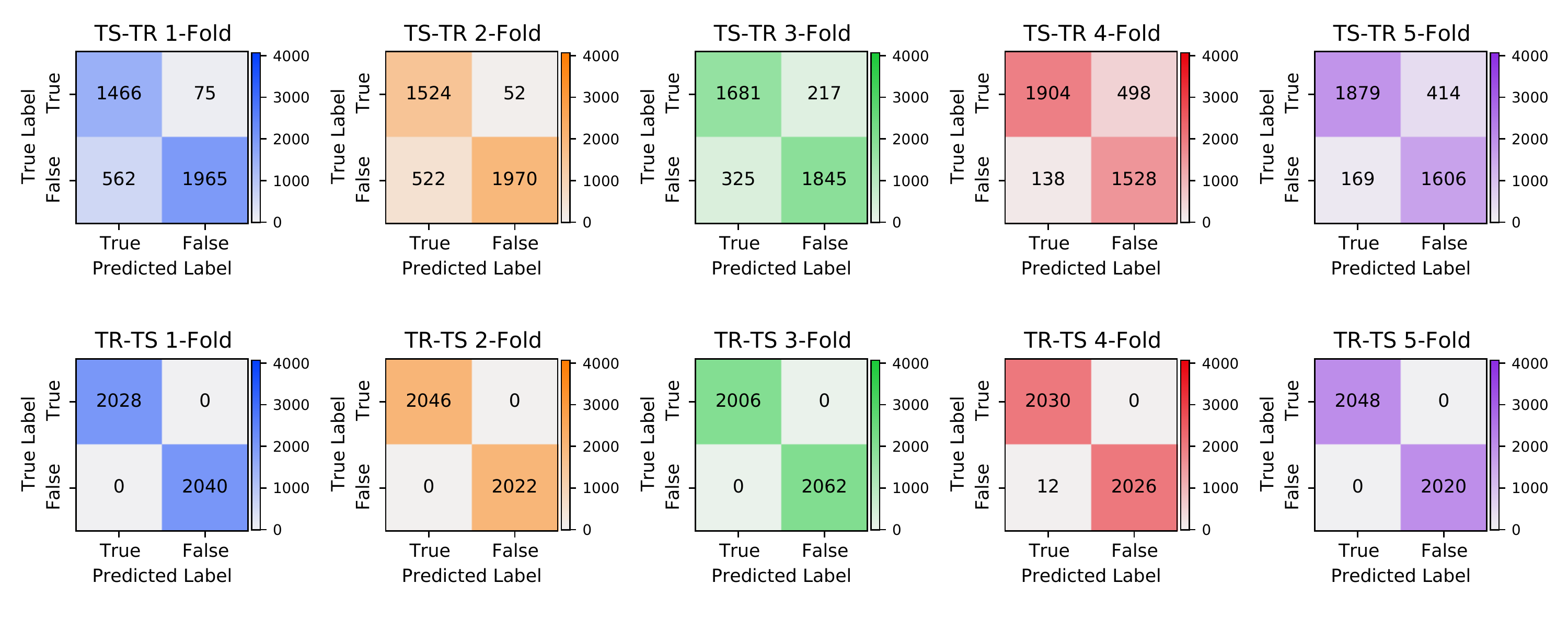}
    \caption{Evaluating of Adversarial model via SVM Confusion Matrices.}
    \label{fig_confusion_matrix}
\end{figure}

\begin{figure}[!htb]
    \centering
    \includegraphics[width=1\linewidth]{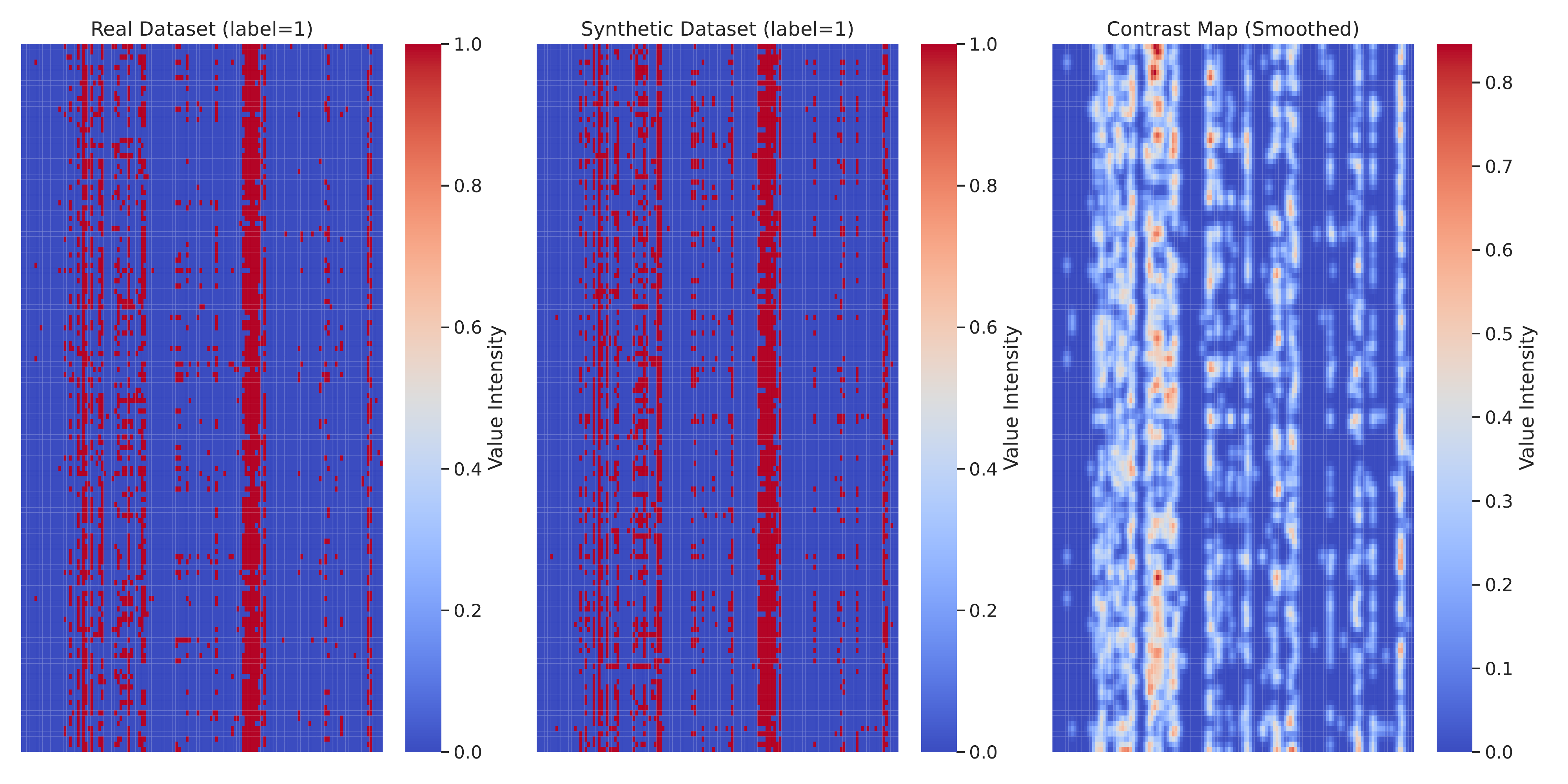}
    \caption{Comparative heat map.}
    \label{fig_heat_map}
\end{figure}

\begin{figure}[!htb]
    \centering
    \includegraphics[width=1\linewidth]{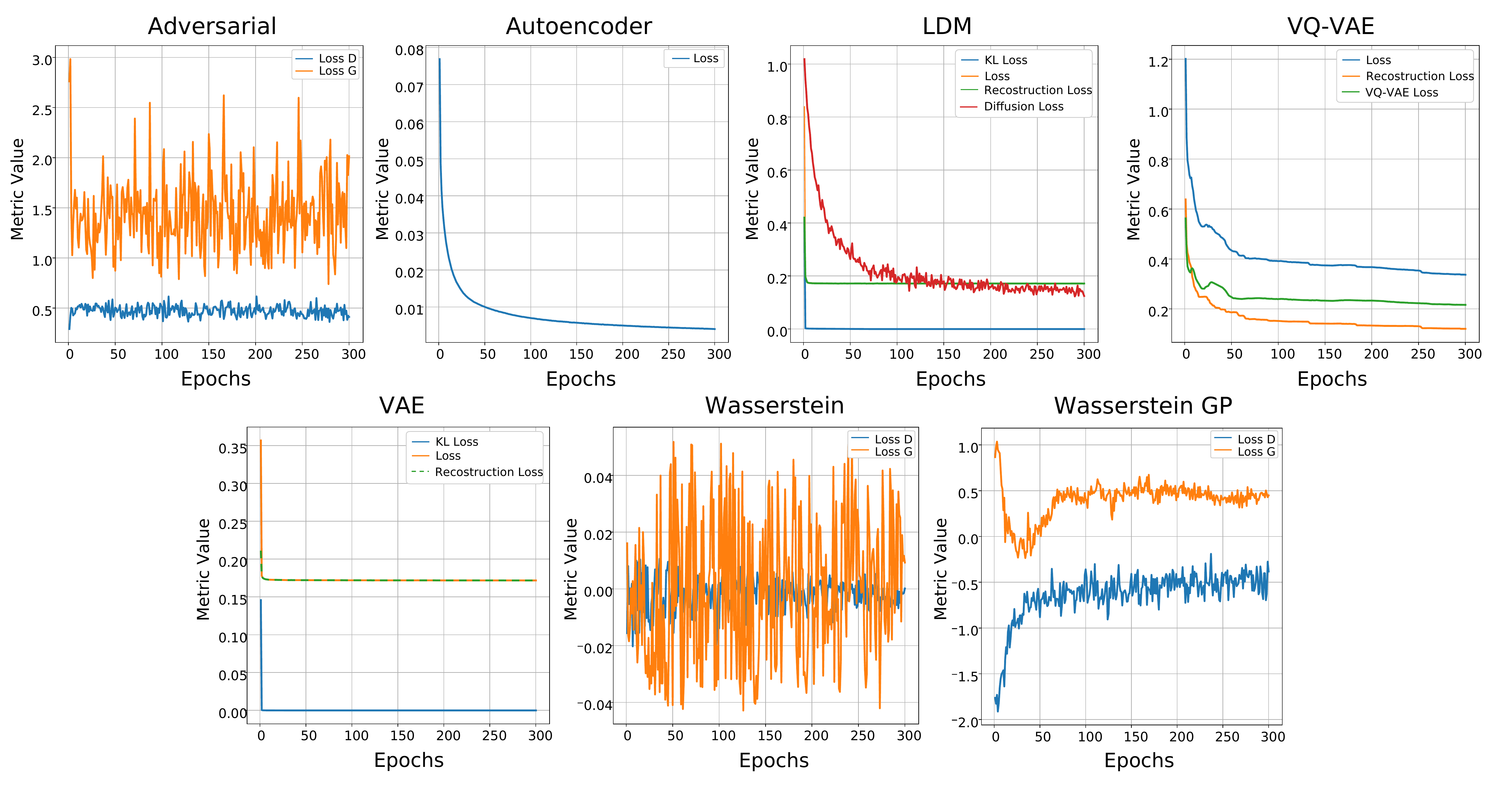}
    \caption{Training loss curves of the models.}
    \label{fig_training_curve}
\end{figure}

Figures~\ref{fig_confusion_matrix}, \ref{fig_heat_map} and \ref{fig_training_curve} encapsulate our evaluation pipeline’s key results. The aggregated confusion matrices in Figure~\ref{fig_confusion_matrix} summarize binary classification performance (TP, FN, TN, FP) under both TR‑TS and TS‑TR protocols. Figure~\ref{fig_heat_map} compares real versus synthetic feature means, with an overlaid plot highlighting divergences to assess generation fidelity. Lastly, Figure~\ref{fig_training_curve} traces each model’s training stability: WGAN/WGAN‑GP generator and discriminator losses, VQ‑VAE’s total, reconstruction and quantization losses, and a single reconstruction curve for the autoencoder. Collectively, these visuals demonstrate the synthetic data’s quality and the robustness of our training workflows.

\end{document}